\documentclass[12pt,twocolumn]{elsart}
\usepackage{graphicx}

\begin{document}

\begin{frontmatter}
\title{The ATLAS Pixel Detector}
\runauthor{J\"{o}rn Grosse-Knetter for ATLAS Pixel}
\author[BN]{J\"{o}rn Grosse-Knetter},
\ead{jgrosse@physik.uni-bonn.de}
\author{on behalf of the ATLAS Pixel collaboration}
\address[BN]{Physikalisches Institut, Universit\"{a}t Bonn, 
  Nussallee 12, D-53115 Bonn, Germany}

\begin{abstract}
The ATLAS Pixel Detector is the innermost layer of the ATLAS tracking 
system and will contribute significantly to  the ATLAS track and vertex 
reconstruction. The detector consists of identical sensor-chip-hybrid modules,
arranged in three barrels in the centre and three disks on either side
for the forward region.

The position of the Pixel Detector near the interaction point requires 
excellent radiation hardness, mechanical and thermal robustness, good 
long-term stability, all combined with a low material budget. 
The detector layout, results from final prototyping and the status 
of production are presented.
\end{abstract}

\end{frontmatter}

\section{Introduction}

The ATLAS Inner Detector~\cite{IDTDR}
is designed for precision tracking of charged 
particles with 40~MHz bunch crossing identification. It
combines tracking straw tubes in the outer transition-radiation tracker (TRT)
and microstrip detectors of the semiconductor tracker (SCT) 
in the middle with the Pixel Detector, the crucial part for
vertex reconstruction, as the innermost
component.

The Pixel Detector~\cite{PDTDR} is subdivided into
three barrel layers in its centre, one of them around the beam pipe ($r=5$~cm),
and three disks on either side for the forward direction. With a total length of
approx. 1.3~m
this results in a three-hit system for particles with $|\eta|<2.5$.

The main components are approx.~1700 identical sensor-chip-hybrid modules,
corresponding to a total of $8\cdot 10^7$ pixels.
The modules have to be radiation hard to an ATLAS life time dose 
of 50~MRad or $10^{15}$ neutron-equivalent.

\section{Module Layout}

A pixel module consists of a single n-on-n silicon sensor, 
approx.~2$\times$6~cm$^2$ in size. The sensor is
subdivided into 47,268 pixels which are connected individually to 16 
front-end (FE) chips via ``bumps''~\cite{fabian}. These chips are connected to a
module-control chip (MCC)~\cite{MCC} mounted on a kapton-flex-hybrid glued 
onto the back-side of the sensor. 
The MCC communicates with the off-detector electronics via 
opto-links, whereas power is fed into the chips via cables connected 
to the flex-hybrid.

To provide a high space-point resolution of approx.~12$\,\mu$m in azimuth ($r\phi$), 
and approx.~90$\,\mu$m parallel to the LHC beam ($z$),
the sensor is subdivided into
41,984 ``standard'' pixels of 50~$\mu$m in $r\phi$ by 400~$\mu$m in $z$,
and 5284 ``long'' pixels of $50 \times 600$~$\mu$m$^2$.
The long pixels are necessary 
to cover the gaps between adjacent front-end chips. 
The module has 46,080 read-out channels, 
which is smaller 
than the number of pixels because there is a 
200~$\mu$m gap in between
FE chips on opposite sides of the module, 
and to get full coverage the last eight pixels at the gap
must be connected 
to only four channels (``ganged'' pixels).
Thus  on 5\% of the surface the information has a two-fold ambiguity 
that will be resolved off-line.

The FE chips~\cite{fabian} contain 2880 individual charge sensitive 
analogue circuits with a digital read-out
that operates at 40~MHz clock. The analogue part consists 
of a high-gain, fast preamplifier followed
by a DC-coupled second stage and a differential discriminator. 
The threshold of the discriminator
ranges up to 1~fC, its nominal value being 0.5~fC. 
When a hit is detected by the discriminator the 
pixel address 
is provided together with the time over threshold 
(ToT) information which allows reconstruction
of the charge seen by the preamplifier. 

\section{Module Prototypes}

Several ten prototype modules have been built with a first generation of 
radiation-hard chips in $0.25\,\mu$m-technology. 
In order to assure full functionality of the modules in the 
later experiment, measurements
at the production sites, after irradiation, and
in a test beam are performed.

\subsection{Laboratory measurements}

\begin{figure*}[t]
  \centerline{\hbox{ \hspace{.03\columnwidth}
    \includegraphics[width=\columnwidth]{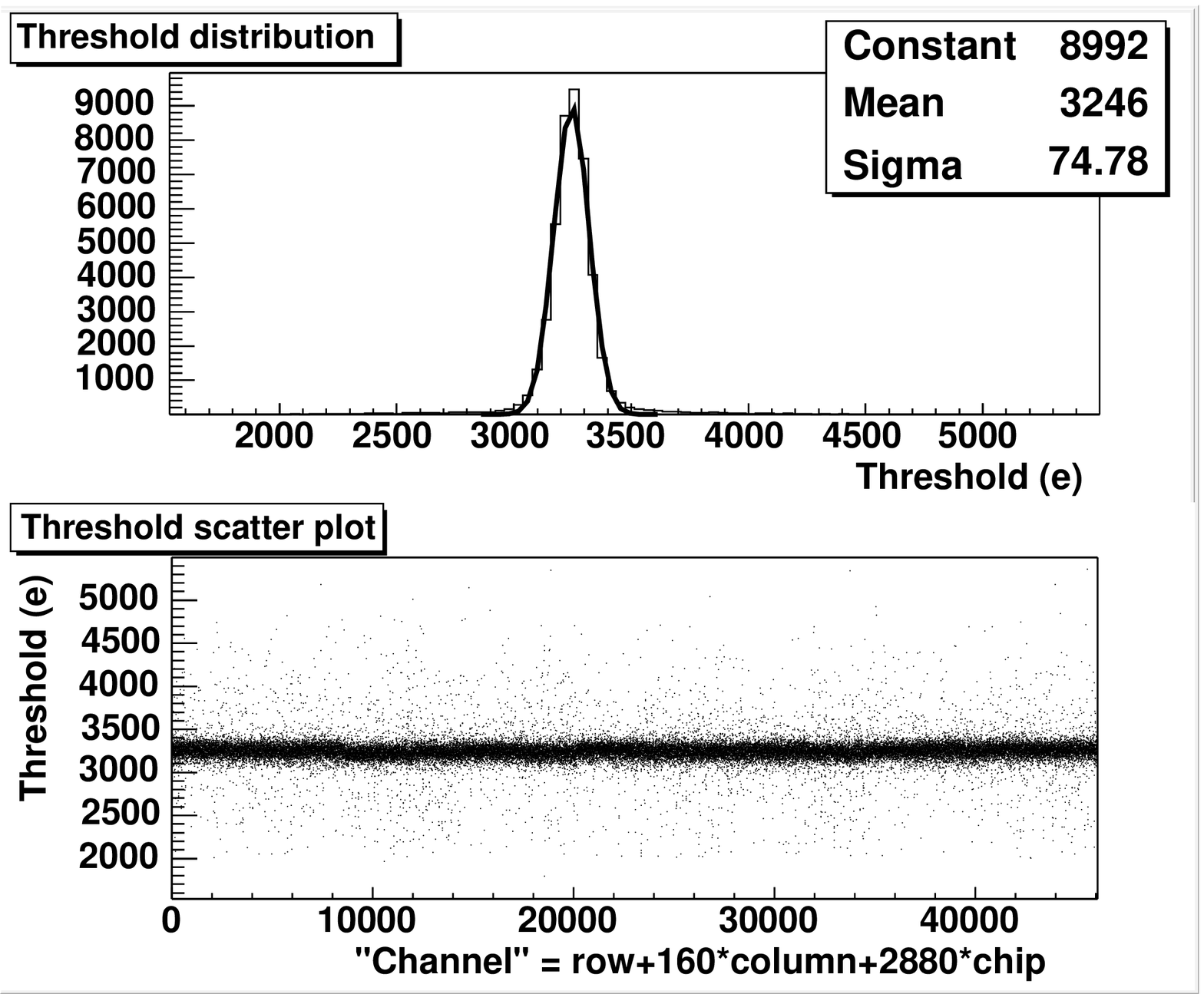}
    \hspace{.05\columnwidth}
    \includegraphics[width=\columnwidth]{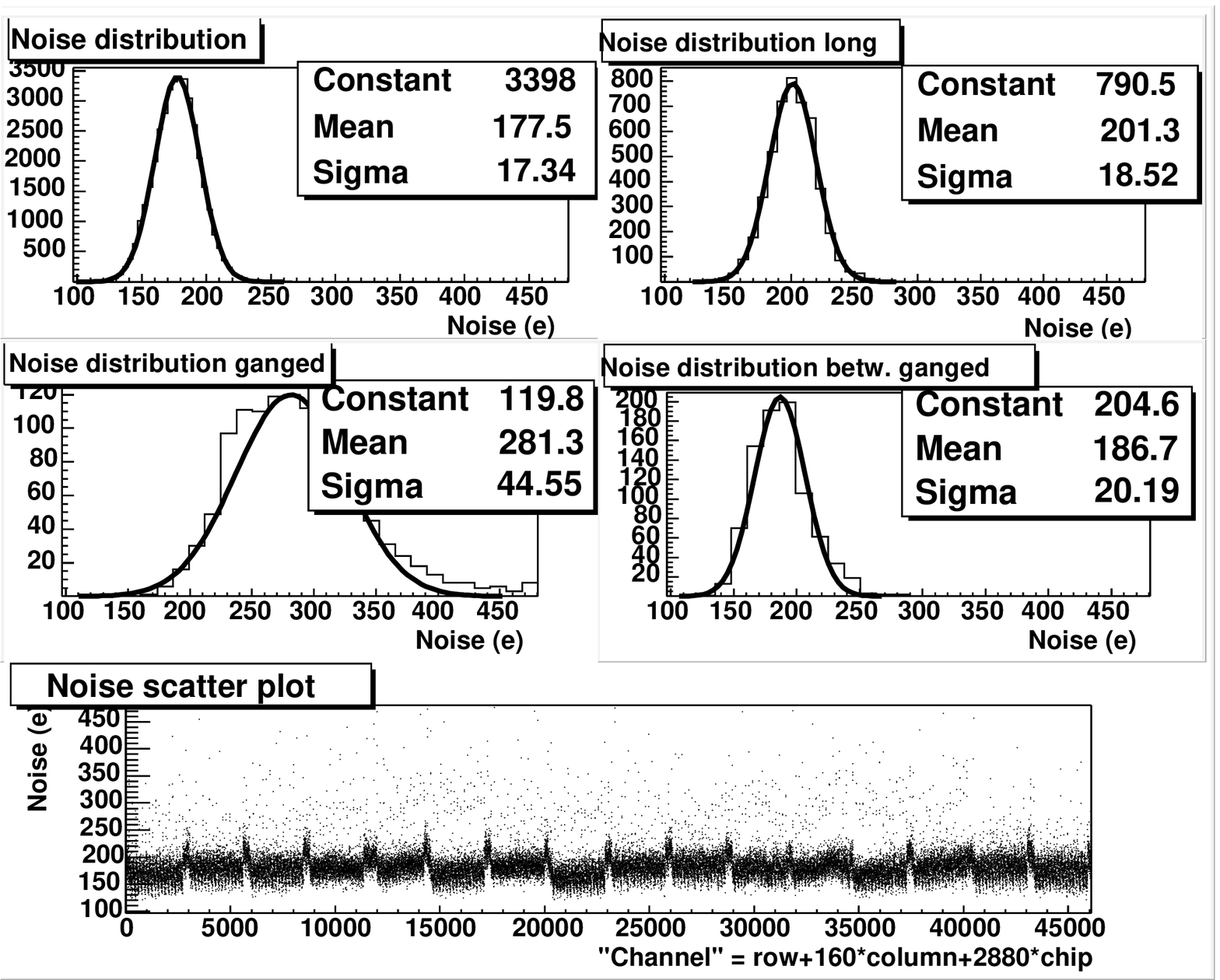}
    }
  }
 \caption{
      Distributions of threshold (left) and noise (right) of a module
      after individual threshold tuning.
    \label{fig:thrnse} }
\end{figure*}

An important test that allows a large range of in-laboratory 
measurements is the threshold scan.
Signals are created with on-module charge injection and 
scanning the number of hits
versus the so injected charge yields the physical value of 
the threshold of the discriminator
and the equivalent noise charge as seen by the preamplifier. 
A set of such scans is used to reduce the
threshold dispersion by adjusting
a DAC-parameter individually for each channel.
The resulting threshold
and noise after threshold tuning is shown in 
figure~\ref{fig:thrnse}.
Typically approx.~100~e threshold dispersion across 
a module and a noise value
of below 200~e for standard pixels is achieved, as is 
needed for good performance.
In a similar fashion, the cross-talk is measured to a few per 
cent for 
standard $50\times 400$~$\mu$m$^2$ pixels. 

A measurement of the timewalk, i.e. the variation in the time 
when the discriminator input goes above threshold, is an 
issue since hits with a low 
deposited charge have an arrival time later than those
with high charges,
in particular for ganged pixels. 

Data taken when illuminating the 
sensor with a radioactive source
allows in-laboratory detection of defective channels.
The source-spectrum reconstructed 
from the ToT-readings
is in agreement with expectations. 

\subsection{Irradiation}~\label{sec:irr}

Some of the prototype modules have been irradiated to a dose
of 50~MRad, approx.~the dose expected after 
10 years of ATLAS operation. The radiation
damage is monitored reading the leakage current individually for
each pixel. The single event upset rate
is measured during irradiation.

\begin{figure*}[t]
  \centerline{\hbox{ \hspace{.03\columnwidth}
    \includegraphics[height=\columnwidth, angle=270]{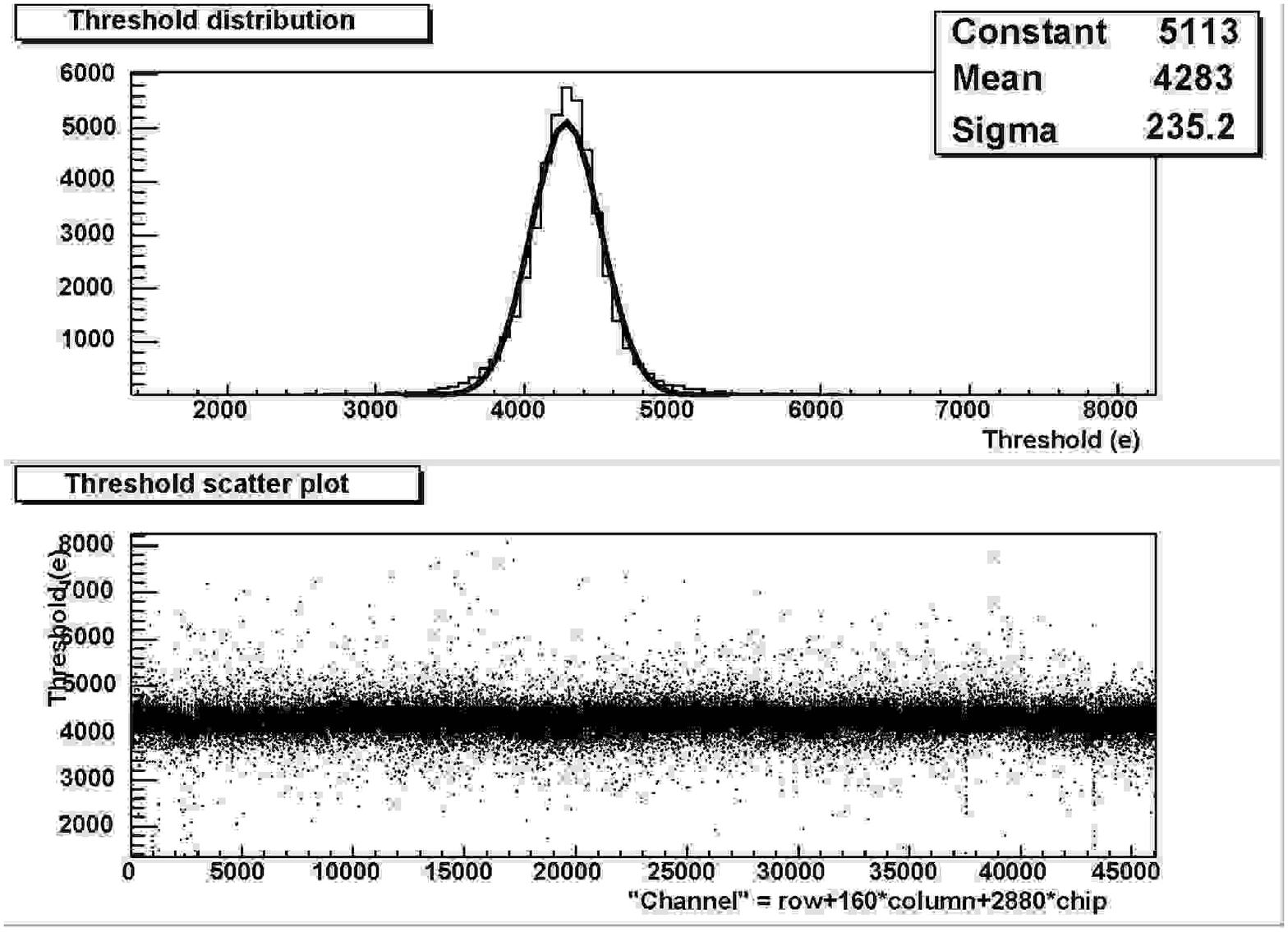}
    \hspace{.05\columnwidth}
    \includegraphics[height=\columnwidth, angle=270]{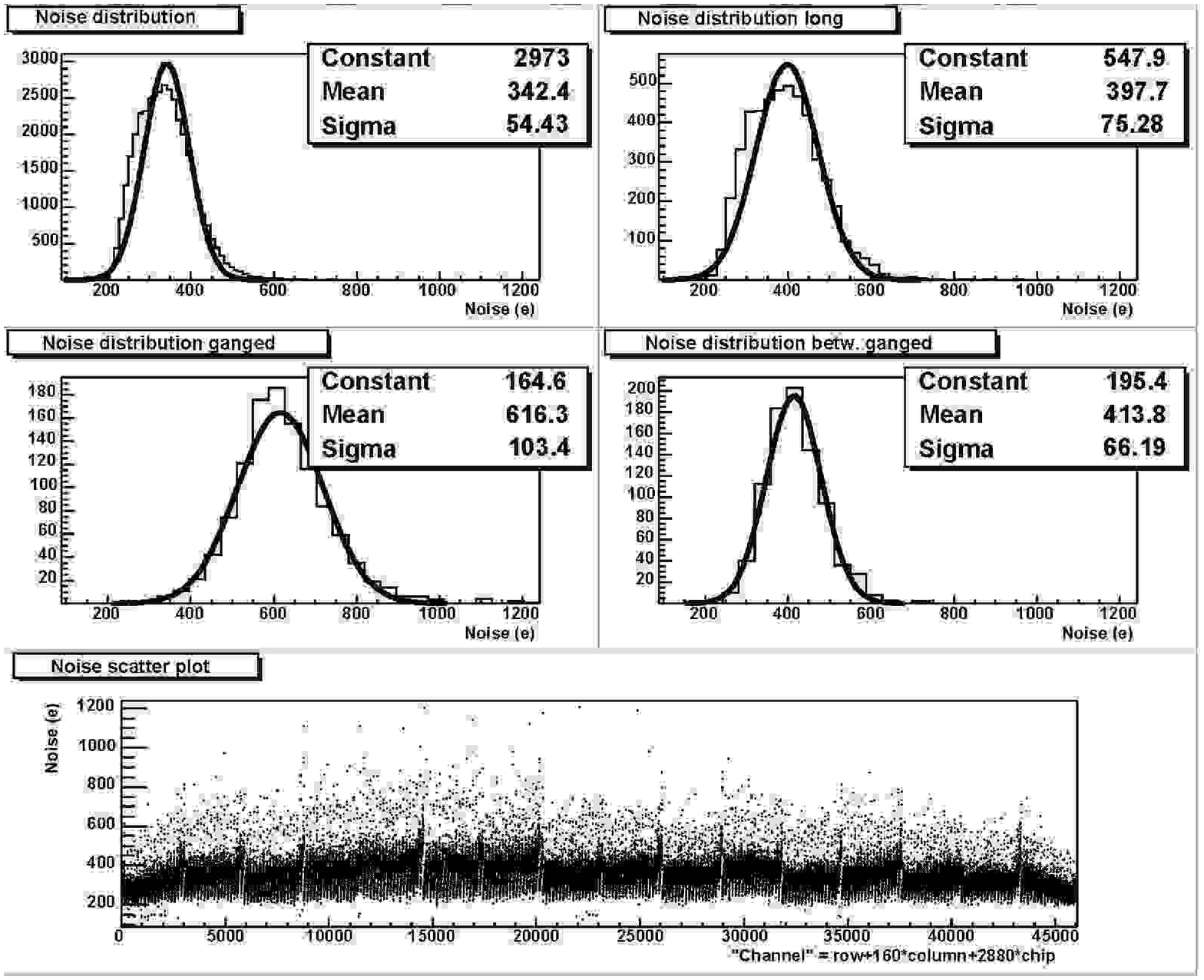}
    }
  }
 \caption{
      Distribution of threshold (left) and noise (right) of a module
      after irradiation with 50~MRad.
    \label{fig:irrtn} }
\end{figure*}

The threshold dispersion and the noise after irradiation as shown in 
figure~\ref{fig:irrtn} is only modestly increased and still well
in agreement with requirements for operation in ATLAS.

\subsection{Test beam measurements}

Tests have been performed in the beam line of the SPS at CERN 
using 180~GeV/c hadrons. 
The setup consists of a beam telescope for the position 
measurement~\cite{BAT}, trigger 
scintillators for timing measurement to 36~ps, and up to four pixel modules.
The number of defective channels is observed to less than $10^{-3}$ and 
for standard $50\times 400$~$\mu$m$^2$ 
pixels the efficiency for normal 
incidence particles is 99.57$\pm$0.15\%. 
The timewalk is measured to values similar
to those from lab tests (see above).

\begin{figure}[b]
    \centerline{\includegraphics[height=.9\columnwidth, angle=270]{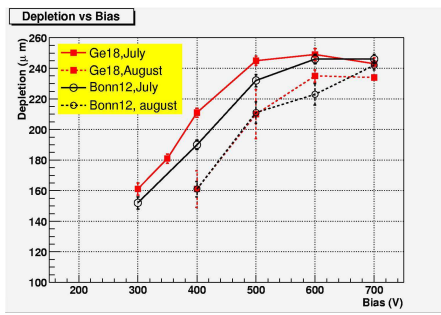}}
 \caption{
        Depletion depth measured in a test beam as a function of
        bias voltage.
    \label{fig:deplV} }
\end{figure}

Modules irradiated as described in section~\ref{sec:irr} have been tested
in the beam line and the bias voltage needed for full depletion is measured
to be between 500 and 600~V, see figure~\ref{fig:deplV}. The
deposited charge is measured via the ToT readings and no significant
changes in the uniformity w.r.t.~unirradiated modules are observed.

\subsection{Next generation of chips}

The results of the prototype modules show that the 
chips fulfil largely the requirements of electrical performance and
radiation hardness and that the production process has a high yield.
Problems such as timewalk and too little SEU tolerance
have been addressed and solved in a new design of FE and 
MCC chips~\cite{fabian}.

\section{Off-detector electronics}

The off-detector readout electronics is designed to process data 
at a rate of up to 100~kHz level-1 triggers. 
The main data-processing component is the ``read-out driver'' (ROD), 
of which first prototypes
have been built to pixel specifications
and are being evaluated.
The first-step event-building and error flagging is done via FPGAs.
The communication 
to the rest of the DAQ-system is run through a 1.6~Gbit/s opto-link.
The communication to modules, online monitoring and calibration runs are
performed with SRAMs and DSPs; their programming is ongoing and
modules have already been configured and operated successfully with a ROD.

\section{System aspects}

\subsection{Support structures}

The mechanics of the system has to guarantee good positional
stability of the modules during operation while the
amount of material has to be kept to a minimum.
At the same time it has to
provide cooling to remove the heat load from the modules
and maintain the sensors at a temperature of -$6^\circ$C to
keep the radiation damage low.

Barrel-modules are glued to ``staves'',
long, flat carbon-structures with cooling pipes embedded.
The staves are mounted inside halfshells, which themselves
are assembled into frames to form the barrel system.

The disks are assembled from carbon-sectors with embedded
cooling covering
1/12 of a wheel. The modules are glued directly to either
side of the disk sectors. 

\subsection{Systemtests}

\begin{figure}[t]
    \centerline{\includegraphics[width=.9\columnwidth]{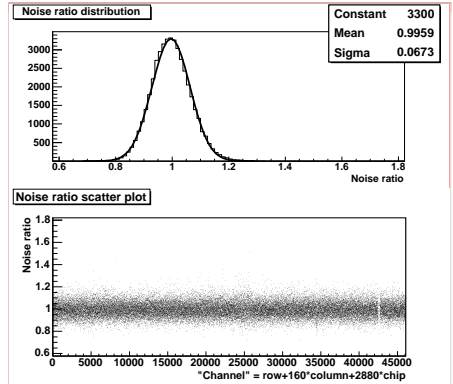}}
 \caption{
        Ratio of noise on one module comparing between simultaneous operation
        with two other modules and individual operation.
    \label{fig:sysnse} }
\end{figure}

First mini-systemtests have been performed
with six modules on a disk sector and
three modules on a barrel-stave.
The noise behaviour on the disks or staves
shows no significant differences
compared to similar measurements with the same 
modules individually, see
figure~\ref{fig:sysnse}.
Larger systemtests are already in preparation and will include
realistic powering and read-out.

\section{Conclusions}

Prototype modules built with a first generation
of radiation hard chips show largely satisfying performance
in laboratory-test, in test beam studies and after irradiation. 
Remaining problems have been solved in a new generation of chips which
is now ready for production. 

Work on the off-detector electronics and the support structures
have been going on in parallel and are well on track. First
systemtest results are promising.

\end{document}